\def\kms{km~s$^{-1}$}

\def\oii{[O~{\sc ii}]}

\def\hi{H~{\sc i}}

\def\nhi{\mbox{$\sc N(\sc H~{\sc I})$}}

\def\feii{Fe~{\sc ii}}

\def\mgi{Mg~{\sc i}}
\def\mgii{Mg~{\sc ii}}

\documentclass[fleqn,usenatbib]{mnras}

\usepackage{newtxtext,newtxmath}
\usepackage[T1]{fontenc}
\usepackage{ae,aecompl}

\usepackage{graphicx}	% Including figure files
\usepackage{amsmath}	% Advanced maths commands
\usepackage{amssymb}	% Extra maths symbols

\title[Spatially Resolved Metal Gas Clouds]{Spatially Resolved Metal Gas Clouds}

\author[C. P\'eroux et al.]{C. P\'eroux$^{1}$\thanks{E-mail:celine.peroux@gmail.com},
H. Rahmani$^{1,2}$,
F. Arrigoni Battaia$^{3}$
and R. Augustin$^{1,3}$
\\
$^{1}$ Aix Marseille Universit\'e, CNRS, LAM (Laboratoire d'Astrophysique de Marseille) UMR 7326, 13388, Marseille, France.\\
$^{2}$ GEPI, Observatoire de Paris, PSL Universit\'e, CNRS,  5 Place Jules Janssen, 92190 Meudon, France.\\
$^{3}$ European Southern Observatory (ESO), Karl-Schwarzschild-Str.2, D-85748 Garching b. M\"unchen, Germany.
}

\date{Accepted 2018 May 18. Received 2018 May 17; in original form 2018 March 5
}

\pubyear{2018}

\begin{document}
\label{firstpage}
\pagerange{\pageref{firstpage}--\pageref{lastpage}}
\maketitle

\vspace{-0.7cm}

\begin{abstract}
We now have mounting evidences that the circumgalactic medium (CGM) of galaxies is polluted with metals processed through stars. The fate of these metals is however still an open question and several findings indicate that they remain poorly mixed. A powerful tool to study the low-density gas of the CGM is offered by absorption lines in quasar spectra, although the information retrieved is limited to 1D along the sightline. We report the serendipitous discovery of two close-by bright z$_{\rm gal}$=1.148 extended galaxies with a fortuitous intervening z$_{\rm abs}$=1.067 foreground absorber. MUSE IFU observations spatially probes kpc-scales in absorption in the plane of the sky over a total area spanning $\sim$30 kpc$^{-2}$. We identify two \oii\ emitters at z$_{\rm abs}$  down to 21 kpc with SFR$\sim$2 M$_{\odot}$/yr. We measure small fractional variations ($<$30\%) in the equivalent widths of \feii\ and \mgii\ cold gas absorbers on coherence scales of 8kpc but stronger variation on larger scales (25kpc). We compute the corresponding cloud gas mass $<$2$\times$10$^{9}$M$_{\odot}$. Our results indicate a good efficiency of the metal mixing on kpc-scales in the CGM of a typical z$\sim$1 galaxy. This study show-cases new prospects for mapping the distribution and sizes of metal clouds observed in absorption against extended background sources with 3D spectroscopy. 
\end{abstract}


%%%%%%%%%%%%%%%%%%%%%%%%%%%%%%%%%%%%%%%%%%%%%%%%%%

%%%%%%%%%%%%%%%%% BODY OF PAPER %%%%%%%%%%%%%%%%%%

\section{Introduction}

Baryons from the cosmic web are known to accrete efficiently onto galaxies. This mechanism sustains violent episodes of star formation which power outflows extending out to the surrounding circumgalactic medium (CGM), and even reaching the larger scales of the intergalactic medium or IGM \citep{aguirre01,oppenheimer06}. The metals carried by the outflows will either rain back on to galaxies or get mixed into the IGM. Indeed, observations of the IGM indicate significant quantities of metals at all redshifts \citep{pettini03, ryanweber09, dodorico13, shull14a}. However, the mixing could remain incomplete \citep[e.g.][]{dedikov04}. Specifically, \citet{schaye07} show that ionised metal clouds are compact (typical scales of 100pc) and are short-lived. Once in the IGM they expand until they reach pressure equilibrium with their environment but remain poorly mixed on scales of $\sim$1kpc or smaller.
More recently, \citet{churchill15} find that in the CGM of a simulated dwarf galaxy, low ionization gas arises 
cloud structures of scales of order $\sim$ 3kpc. High ionization gas however lies in multiple extended structures spread over 100 kpc and due to complex velocity fields, highly separated structures give rise to absorption at similar velocities \citep[see also][]{bird15}. These authors predict a mismatch between the smoothing scales of \hi\ and high-ionisation metals which has yet to be witnessed with observations. These findings question how traditional absorption lines studies recover the true gas properties. Based on hydrodynamical equilibrium arguments, \citet{mccourt18} further suggest that the CGM of galaxies cools via "shattering", resulting in a high covering fraction of pc-scale, photoionised cloudlets. 

Important issues thus remain unsolved: On which scales are metals mixed? How does it vary with environment (i.e. IGM vs. CGM)? Do low and high-ionisation ions have different coherence scales? Can we find direct observational evidence of shattering on pc-scales? Where are the high-metallicity, intergalactic gas clouds? If the \hi\ absorption and high-ionisation metal absorption gas arise in distinct physical gas structures, the observational techniques employed to infer metallicities and the total mass of the warm-hot CGM gas would be challenged \citep{tumlinson11, werk14}. The pockets of metal-rich material have also profound implications on metal-cooling efficiency and in turn galaxy formation. Our understanding of these phases of the gas and their metallicities has so far been limited by the lack of observational constraints. 

A powerful tool to study this low-density gas is offered by absorption lines in quasar spectra. In these quasar absorbers, the minimum gas density that can be detected is set by the brightness of the background source and thus the detection efficiency is independent of redshift. High-quality quasar absorption spectra have produced a wealth of information regarding the distribution of heavy elements \citep[e.g.][]{kulkarni05, quiret16}. However, the brightest background sources (quasars and gamma-ray bursts) are point-sources so that the observer is limited to the information gained along the line-of-sight. The metallicity we typically infer from absorption studies is then not determined by the abundances of heavy elements on the size of the metal concentrations, but by the metallicity smoothed over the size of the HI absorber, which is typically 100 kpc \citep{bechtold94, schaye03, schaye05}. 
On smaller scales the distribution of metals is essentially unknown. To remedy this, observers have used close quasar pairs \citep{hennawi06, martin10, rubin15, rubin18} as well as multiple images from gravitationally lensed background objects to probe the transverse small-scale coherence along lines-of-sight dozen kpc apart \citep{rauch01, ellison04, lopez07, chen14, rubin17}. 

By using an extended galaxy as background source however, one can directly map the distribution and sizes of the metal absorbers on small scales. By using this set-up, \citet{cooke15} have estimated that indeed high-column density neutral gas can span continuous areas 10$^8$--10$^{10}$ times larger than previously explored in quasar or gamma-ray burst sightlines. \citet{bergeron17} have used intervening absorber lines redshifted on more extended quasar emission lines to probe the spatial covering of the gas clouds. In a remarkable work, \citet{lopez18} reported \mgii\ absorption along a bright lensed arc probing scales of the order 2--4kpc. They find that the strength of the absorber decreases with radius from the emitting galaxy as expected from the quasar absorber population. However the physical properties derived from the observations of lensed systems rely heavily on the lensed model to compute the magnification factors and survey different physical area in the image plane, thus probing inhomogeneous flux levels.

Here, we report the serendipitous discovery of two bright z$_{\rm gal}$=1.148 extended galaxies with a fortuitous intervening foreground absorber at z$_{\rm abs}$=1.067 along their sightlines. The manuscript is organised as follows: Section 2 provides the observational set-up and details of the system lay-out. Section 3 presents the constraints on the physical conditions of the metal gas clouds. Finally, in section 4, we review the impact of these findings in the broader context of the metal mixing in the CGM of galaxies. Throughout this paper we adopt an $H_{0}=70$~\kms~Mpc$^{-1}$, $\Omega_{\rm M}=0.3$, and $\Omega_{\rm \Lambda}=0.7$ cosmology. At the redshift of the absorber (z$_{\rm abs}$=1.067), 1" corresponds to 8.2 kpc.

%%%%%%%%%%%%%%%%%%%%%%%%%%%%%%%%%%%%
%%%%%%%%%%%%%%%%%%%%%%%%%%%%%%%%%%%%
%%%%%%%%%%%%%%%%%%%%%%%%%%%%%%%%%%%%
\section{MUSE Observations of a Remarkable System}

\begin{figure}
	\includegraphics[width=0.38\textwidth]{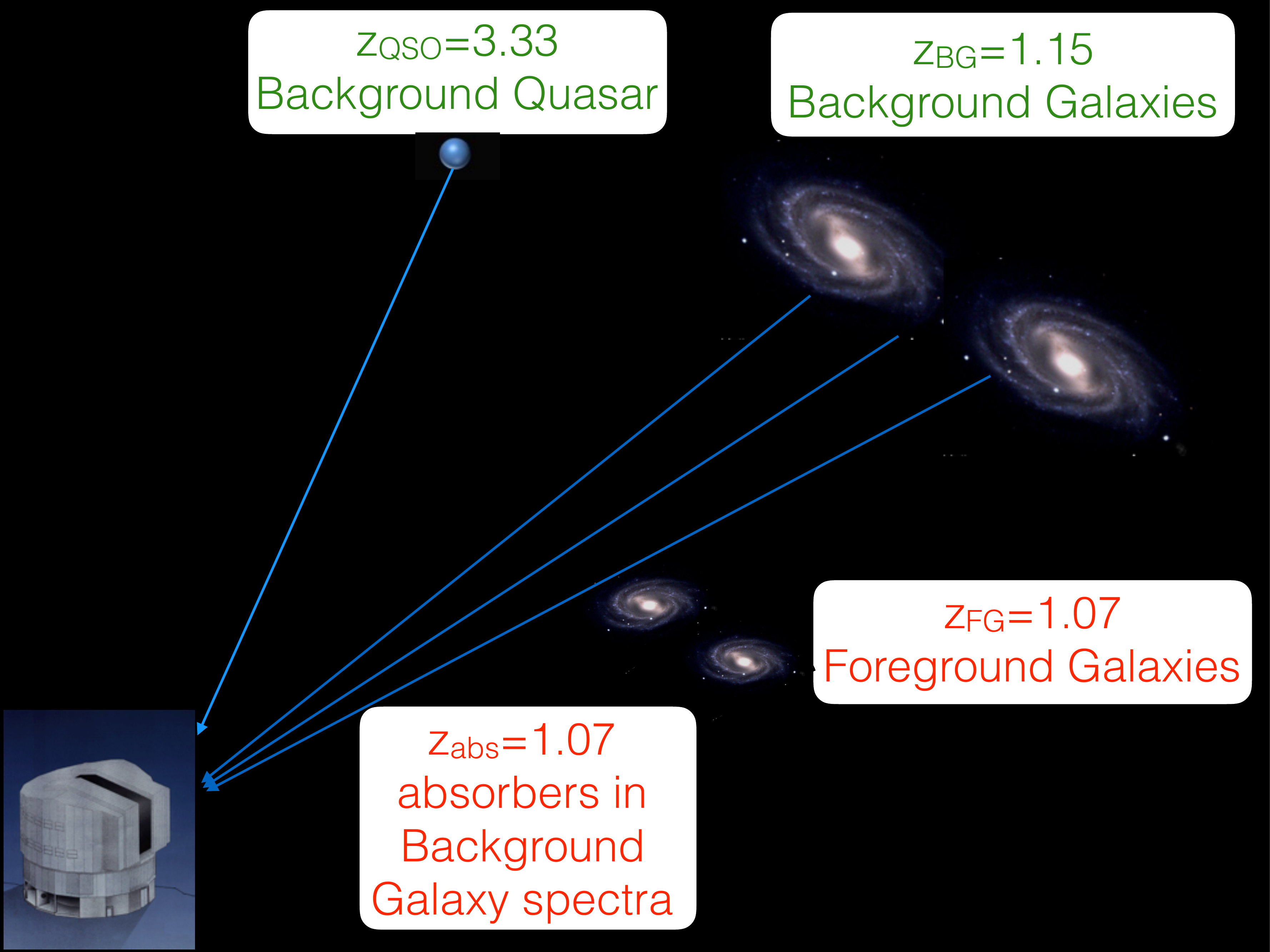}	
  \caption{Sketch of the system lay-out. A bright z$_{\rm qso}$=3.33 quasar was the original target of the observations. In these data, we have serendipitously discovered an intervening z$_{\rm abs}$=1.067 metal absorber in the spectrum of two close-by z$_{\rm gal}$=1.148 extended background galaxies. In addition, two emitting galaxies are observed in \oii\ at the redshift of the absorber.}
  \label{f:sketch}
\end{figure}

MUSE observations of the field of quasar SDSS J0250$-$0757 were undertaken in service mode in natural seeing mode under programme 095.A-0615(A). These observations
are part of the survey
QSO MUSEUM (Quasar Snapshot Observations with MUse:
Search for Extended Ultraviolet eMission; Arrigoni Battaia et al.
in prep.). 
The observations were carried out during UT 2015 September 17 in nominal mode and consisted of three
exposures of 900 s each. The sub-exposures were rotated by  90 degrees to minimise residuals from uneven flat-fielding.
The field of view is 60 " $\times$ 60 ", a 0.2 "/pixel scale and a spectral sampling of 1.25 \AA/pixel covering 4750--9350\AA.

%%DATA REDUCTION
The data were reduced with version v1.6.4 of the ESO MUSE pipeline \citep{weilbacher15} and additional external routines for 
sky subtraction as explained in the following.
Master bias, flat field images and arc lamp exposures were produced based on data taken closest in time to the science frames. The raw science and standard star cubes were then processed correcting the wavelength calibration to a heliocentric reference. We checked the wavelength solution using the known wavelengths of the night-sky OH lines and find it to be accurate within 18 km/s.
The individual exposures were registered using the central quasar to ensure accurate relative astrometry. Finally, the individual exposures were combined into a single data cube. 
The removal of OH emission lines from the night sky was accomplished with an additional purpose-developed code tested in previous work \citep{peroux17}. After selecting sky regions in the field, we created Principle Component Analysis (PCA) components from the spectra which were further applied to the science datacube to remove sky line residuals \citep[trouble]{husemann16}. The seeing of the final combined data was measured from the quasar. The resulting point spread function (PSF) has a full width at half maximum of $0.74$" at the wavelength of the \mgii\ absorber ($\sim$5780 \AA), corresponding to 6 kpc at z$_{\rm abs}$=1.067 or $0.68$" at 7700\AA.

%THE SYST
In this MUSE cube, we report the serendipitous discovery of two nearby z$_{\rm gal}$=1.148 \oii\ emitters, coined BGa and BGb, with bright continua (R mag=21.9 and 23.3 respectively). The spectra of these background galaxies show evidences of a "down-the-barrel" outflow with strong absorptions in \mgii, \mgi\ and \feii\ typical of bright galaxies at z=1 \citep{kornei12,martin12}. 
A more remarkable feature however is the presence of strong intervening absorption lines of \mgii\ $\lambda$$\lambda$ 2796, 2803 and \feii\ $\lambda$$\lambda$$\lambda$ 2382, 2586 and 2600 at z$_{\rm abs}$=1.067. The velocity offset between the background galaxies and the intervening absorber is $\Delta v$ $>$11,000 km/s. In addition, two emitting galaxies (FG$\alpha$ and FG$\beta$) are observed in \oii\ at the redshift of the absorber with angular separations of 21 and 63 kpc respectively (measured from the mean of BGa and BGb centroids), well within their CGM regions.  They have R mag=25.7 and 24.5 respectively. The detection limit at this redshift translates into SFR$>0.2$M$_{\odot}$/yr. Figure~\ref{f:sketch} sketches the system lay-out.

\begin{table}
\begin{center}
\caption{ Physical properties of the foreground galaxies FG$\alpha$ and FG$\beta$. 
"b" the impact parameter in kpc, "incl." the inclination and "PA" the position angle (East of North).The offset values are with respect to the sky position of the BGa+BGb background galaxies. The error on the redshift estimates are 0.0001. }
\begin{tabular}{ccccccc}
\hline\hline
Gal.		   &b &$z_{\rm gal}$  &F(\oii) &SFR &incl. &PA\\
&[kpc]&&[erg/s/cm$^2$] &[M$_{\odot}$/yr] &[deg] &[deg]\\
\hline
FG$\alpha$		&21	&1.0677	&3.8$\pm$0.4$\times$10$^{-17}$	&1.8$\pm$0.7 &45$\pm$4 &80$\pm$10\\
FG$\beta$		&63	&1.0677	&4.1$\pm$0.4$\times$10$^{-17}$	&1.9$\pm$0.8 &54$\pm$2 &74$\pm$3\\
\hline\hline 				       			 	 
\end{tabular}			       			 	 
\end{center}			       			 	 
\label{t:em}
\end{table}

\begin{figure*}
	\includegraphics[width=0.55\textwidth]{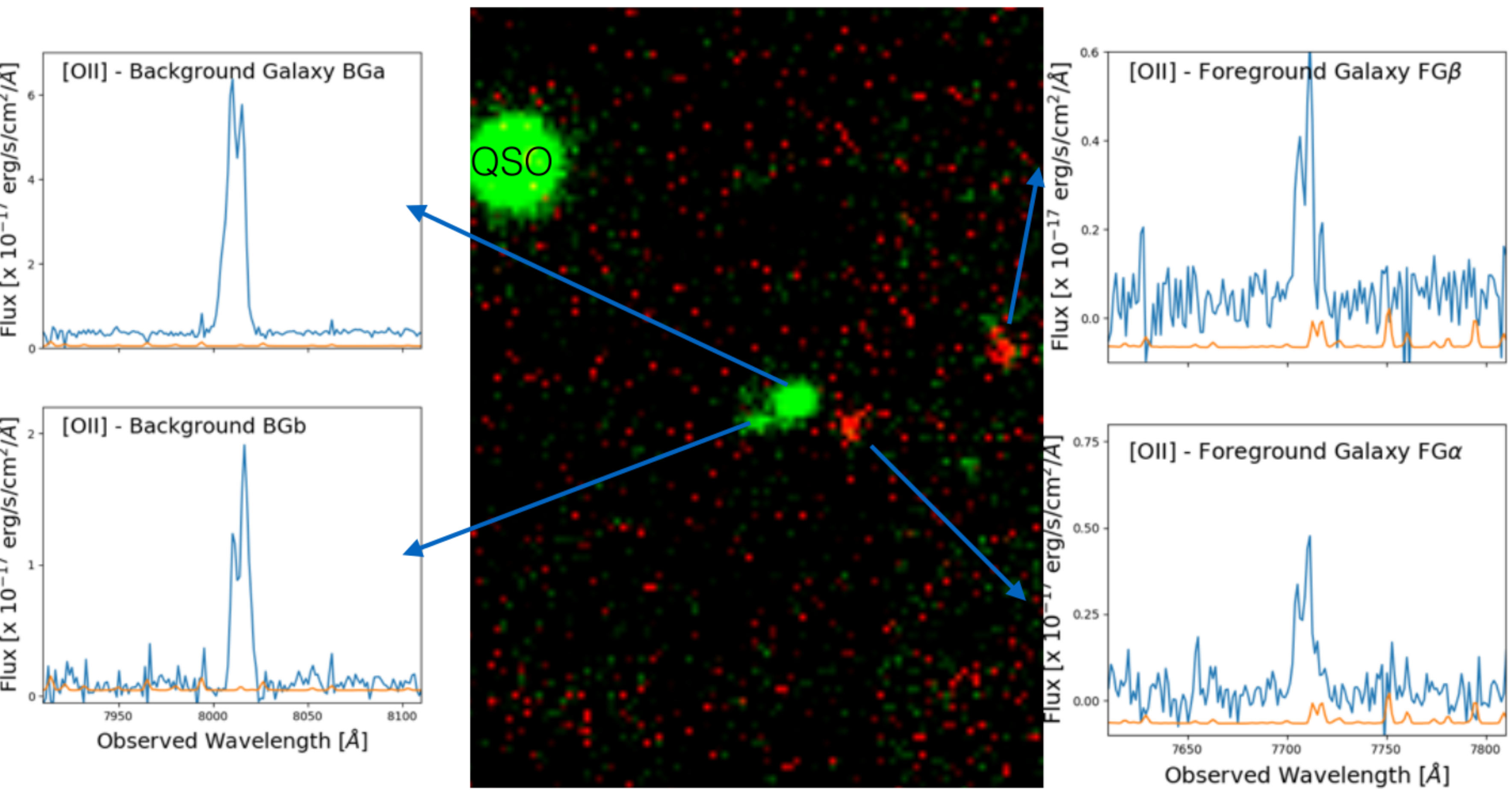}	
  \caption{MUSE observations of the system. Green colours indicate continuum detected objects, while red colours correspond to a pseudo narrow-band filter (7700--7720 \AA) around \oii\ emission at z$_{\rm abs}$=1.067. 
  The spectra on the left show the \oii\ emission lines of the background galaxies BGa and BGb at z$_{\rm gal}$=1.148. The right panel shows the \oii\ emission lines of the foreground galaxies lying at the intervening absorber redshift z$_{\rm abs}$=1.067 (FG$\alpha$ and FG$\beta$). An arbitrarily-scaled sky spectrum is shown in orange.}
  \label{f:em}
\end{figure*}

\begin{table*}
\begin{center}
\caption{Spatial variation of the metal absorber physical properties. $\delta$ is the angular distance in arcsec and "b" the impact parameter in kpc. The offset values and velocity shifts are measured with reference to the sky position and systemic redshift ($z_{\rm fg}$=1.0677) of the \oii\ emitter with the lowest impact parameter, FG$\alpha$. The redshift measurements are the mean of the \feii\ 2600, \mgii\ 2796, 2803 absorption lines measurements, but for BGa4 where only the \mgii\ doublet members are used. Non-detections are quoted as 3-$\sigma$ upper limits.}
\begin{tabular}{lcccccccccc}
\hline\hline
Galaxies		  &$\delta$ &b  &SNR &$z_{\rm abs}$ & velocity &EW$_{\rm FeII~2382}$ &EW$_{\rm FeII~2586}$ &EW$_{\rm FeII~2600}$ &EW$_{\rm MgII~2796}$ &EW$_{\rm MgII~2803}$  \\
&["]&[kpc]&at \mgii&&[km/s] &[\AA] &[\AA] &[\AA] &[\AA] &[\AA]\\
\hline

BGa			&2.0	&16	&8.1	&...					&...				&0.8$\pm$0.2			&0.4$\pm$0.2		&0.9$\pm$0.2		&2.1$\pm$0.3		&1.5$\pm$0.2	\\
...BGa1		&2.3	&19	&4.3	&1.0679$\pm$0.0001	&$+$30$\pm$18	&$<$0.8				&$<$0.8			&1.7$\pm$0.4		&1.9$\pm$0.3  		&1.6$\pm$0.4	\\
...BGa2		&0.6	&5	&4.9	&1.0679$\pm$0.0001	&$+$1$\pm$18		&$<$0.7				&0.8$\pm$0.4		&1.3$\pm$0.5		&2.3$\pm$0.6		&1.8$\pm$0.5	\\
...BGa3		&1.1	&9	&5.4	&1.0675$\pm$0.0001	&$-$30$\pm$18	&$<$0.6				&0.5$\pm$0.3		&1.5$\pm$0.4		&2.7$\pm$0.4		&1.1$\pm$0.4	\\
...BGa4		&2.4	&20	&3.8	&1.0675$\pm$0.0001	&$-$29$\pm$18	&$<$0.9				&$<$0.9			&$<$0.9			&1.9$\pm$0.4		&1.5$\pm$0.4	\\
BGb			&3.1	&25	&2.5	&...					&...				&$<$1.4				&$<$1.4			&$<$1.4			&$<$1.4	 		&$<$1.4		\\
\hline\hline 				       			 	 
\end{tabular}			       			 	 
\end{center}			       			 	 
 \label{t:abs}
 \end{table*}			       			 	 

We model with Galfit the Sersic profile of the background galaxies determining half-light radii, deconvolved from the seeing, of R$_e$=0.28$\pm$0.6" (BGa) and 0.24$\pm$0.05" (BGb), corresponding at the redshift of the absorber to $continuous$ areas of $\sim$17 and 12 kpc$^2$ respectively. This enables us not only to probe the metal cloud over a total area of $\sim$30 kpc$^2$ but also on scales of 25kpc which is the distance in between the galaxies. To extract the spectrum of each galaxy from the MUSE cube, we used MUSE $mpdaf$ v2.5\footnote{http://mpdaf.readthedocs.io/en/latest/index.html} \citep{piqueras17}. We identified pixels associated with each object by running $Sextractor$\footnote{https://www.astromatic.net/software/sextractor} on the 2D white light image. We then extracted the 1D spectrum by integrating the flux of the pixels associated with the objects in each wavelength plane. Because the two background galaxies (separated by 1.5" on the sky) are barely resolved in the MUSE observations and BGb has a faint continuum, we defined a level where the pixels in between the two objects has lower flux values and extract the spectrum of each object from pixels on either side of this threshold. The \oii\ emission of the resulting spectra are shown in Figure~\ref{f:em}. 

The physical properties of the foreground galaxies FG$\alpha$ and FG$\beta$ are summarised in Table 1.
The low impact parameter galaxy (FG$\alpha$) is situated 21kpc away from the background galaxies (BGa+BGb), although we cannot exclude that some absorption components are related to the CGM region of galaxy FG$\beta$.
We measure the \oii\ emission fluxes from a Gaussian fit and derive the SFR estimates, uncorrected for dust  extinction, using the prescription of \citet{kewley04}. The objects have a SFR of a few solar masses per year typical of absorbing galaxies observed at these redshifts \citep{peroux11a, rahmani16}.

%%%%%%%%%%%%%%%%%%%%%%%%%%%%%%%%%%%%
%%%%%%%%%%%%%%%%%%%%%%%%%%%%%%%%%%%%
\section{Properties of Metal Gas Clouds on kpc scales}

\begin{figure}
	\includegraphics[width=0.55\textwidth]{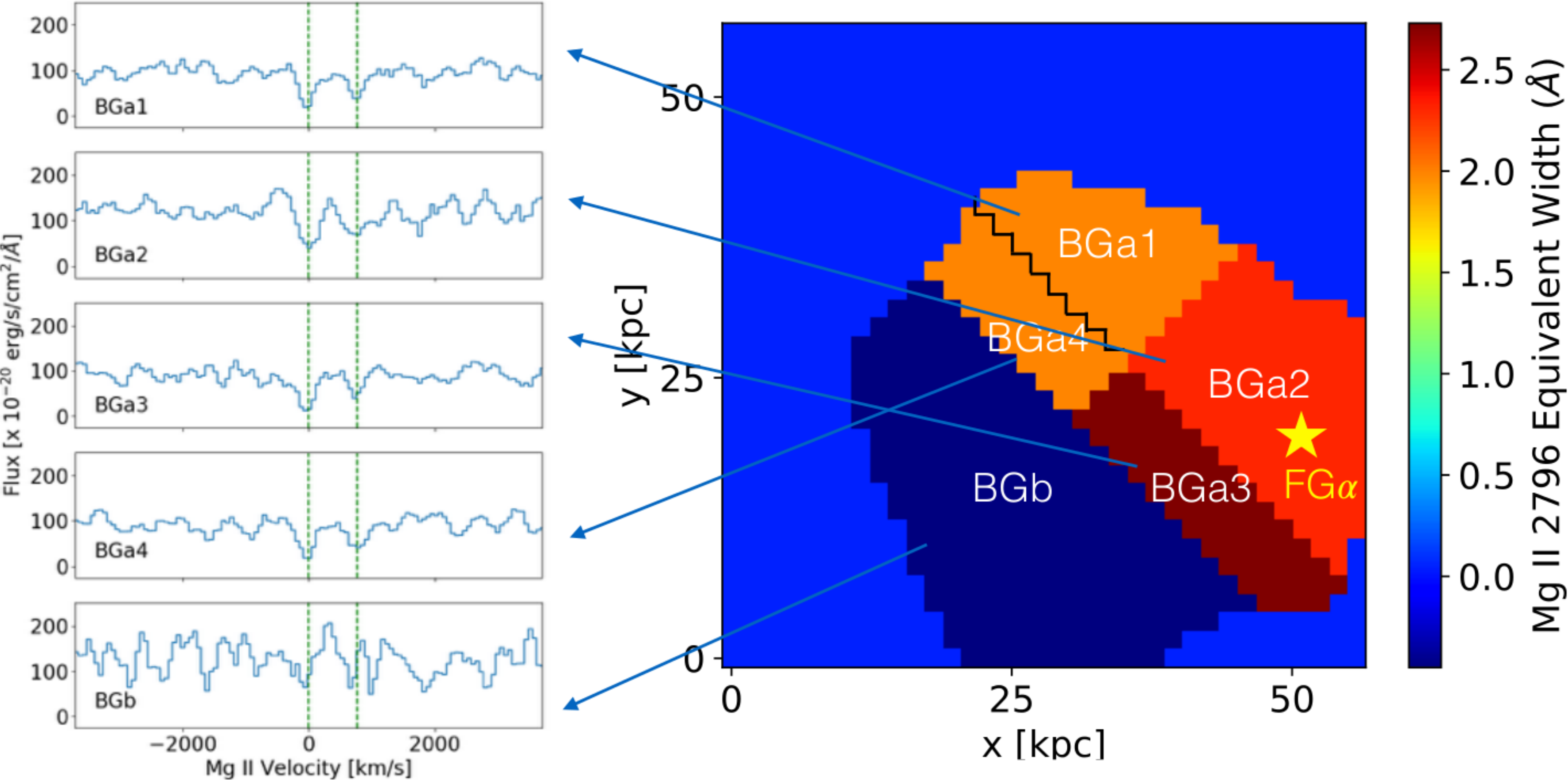}	
  \caption{Physical properties of the metal absorber. The colour map indicates the rest-frame equivalent width of \mgii\ $\lambda 2796$ in units of \AA\ for each region (see text for definition). The corresponding absorbing spectra are also shown in velocity space. The \feii\ lines and \mgii\ doublet are not detected against the background galaxy BGb down to EW$<$1.4 \AA, thus providing an upper limit on the size of the cold metal cloud of R$<$25kpc.}
  \label{f:abs}
\end{figure}

We constrain the physical properties of the metal absorbers within the background galaxy BGa as well as the BGb galaxy. The brightest background object (BGa) is divided in 4 regions separated by  $\sim$8kpc (BGa1, BGa2, BGa3 and BGa4). The location of these regions around the central pixel are not unique but driven by SNR considerations to result in spectra with similar continuum fluxes and hence homogeneous absorption detection limits. Additionally, the seeing-limited observations at hand (FWHM=0.74") imply that the spectra in these regions are partially convolved. The masks used for each of these regions are shown in Figure~\ref{f:abs}. Thanks to the remarkable combination of its high-sensitivity and IFU capabilities, MUSE observations of this system allow us to resolve 4 continuous regions within BGa as well as the background galaxy BGb thus probing sightlines separated by 1" to 3" (8 to 25 kpc at z$_{\rm abs}$). The offset values are measured with reference to the sky position and systemic redshift ($z_{\rm abs}$=1.06768) of the \oii\ emitter with the lowest impact parameter, FG$\alpha$. These values are listed in Table 2.

We measure large rest equivalent widths of \feii\ and \mgii\ metal lines \citep{nielsen13}. With the exception of the absorber against BGa3, the ratios between the measured EWs of \mgii\ lines are $\sim$1.2 indicating line saturation. The limited quality of the data however precludes studies of potential partial coverage.
Non-detections are quoted as 3-$\sigma$ upper limits. 
The \feii\ lines and \mgii\ doublet are not detected against the background galaxy BGb (25 kpc-away from BGa) down to  a significant limit of EW$<$1.4\AA. We note that there are no indications of these metal absorption lines down to EW$<$0.7\AA\ in the bright quasar 14.3" away ($\sim$100kpc at z$_{\rm abs}$). Figure~\ref{f:abs} summarises these findings. The colour map indicates the EW values of each region. The corresponding absorbing spectra are also shown. 

Following \citet{ellison04} and \citet{rubin17}, we calculate the fractional difference in EW values with respect to BGa1. The fractional differences range from 10--20\% (\feii\ $\lambda$ 2600)  to 0--30\% (\mgii\ $\lambda \lambda$ 2796, 2803) in regions where the metal lines are detected. Thus, we report only small variations ($<$30\%) on scales of $\sim$8kpc (i.e. the inter-regions separation against BGa). However, the non-detection of \mgii\ $\lambda$ 2796 against BGb (25 kpc-away) is significant and indicates larger variations ($>$30\%) on this scale. Therefore, while the data at hand show no indication of significant variations on coherence scales of 8kpc, our findings reveal that the metals traced by cold \mgii\ absorptions are inhomogeneously distributed on scales smaller than 25 kpc.

We further compute velocity shifts of the absorber profiles with reference to the systemic redshift of galaxy FG$\alpha$. We measure small shifts of the order 30$\pm$18 km/s with a possible red and blue component. From a fit to the MUSE while light image, we derive the inclination and position angle  of the foreground galaxies (Table 1). The compactness of FG$\alpha$ precludes detailed kinematic analysis \citep[see][]{peroux17} but hints at small rotation velocities. The velocity shear observed in absorption could be the signature of the rotation of a gaseous disk extending from the nearby inclined galaxy FG$\alpha$ (b=21kpc). The small velocities measured could also be produced by turbulent motions over an area containing several clouds of gas. Given the limited spatial resolution of these seeing-limited observations, we cannot disentangle which of these scenarios is at play. 

The metal cloud size estimate is further combined with our knowledge of the density to constrain the cloud gas mass. We 
used the BGa \mgii\ $\lambda 2796$ equivalent width relation with \hi\ column density prescription of \citet{menard09} to estimate the extended gas column density on the sightline of these two background galaxies.
We derive a neutral gas column density of \nhi=1.1$\times$10$^{20}$~cm$^{-2}$. For simplicity we assume the inhomogeneities on 25kpc-scales are due to a spherical cloud, even though the current observations do not rule out asymmetrical geometries (e.g. filament). A conservative circular effective radius $<$25kpc thus corresponds to an area of $<$1.9$\times$10$^{46}$ cm$^{2}$. Assuming a covering factor of unity, we derive a cold metal mass of $<$2$\times$10$^{9}$ M$_{\odot}$. Our data cannot exclude multiple clouds with significantly smaller masses \cite[see e.g.][]{arrigoni15} thus leading to the estimated upper limit. On the other hand, if the cold gas were to arise from a structure centered towards the opposite direction of BGb, the total mass could be higher. Only high spatial resolution observations of a sample of such systems will be able to address these issues. Yet, to our knowledge, these measurements are the first direct estimates of the mass of metal cold gas clouds.

%%%%%%%%%%%%%%%%%%%%%%%%%%%%%%%%%%%%
%%%%%%%%%%%%%%%%%%%%%%%%%%%%%%%%%%%%
%%%%%%%%%%%%%%%%%%%%%%%%%%%%%%%%%%%%
\section{Discussion}
Low-ionisation ions such as \mgii\ and \feii\ are typical tracers of photo-ionised cold gas in galaxies with temperature T $\sim$ 10$^4$ K \citep{bergeron86, charlton03}. They are believed to probe a wide range of neutral hydrogen column densities of \nhi\ $\sim$ 10$^{16}$-10$^{22}$ cm$^{-2}$ \citep{ellison09} within a few hundred kpc of their host galaxies. Simulating this cold phase of the gas has proven challenging because of the complexity of the physics involved and because it requires sub-grid modelling to capture this unresolved physics. However, it is essential to attempt to model the cold gas in the CGM to be able to disentangle different scenarios and velocity signatures for the absorbers (e.g. disks versus turbulent motion of gas). In "zoom-in" simulations (30 h$^{-1}$ pc resolution at z=0), \citet{churchill15} describe these absorbers as clouds, i.e. spatially contiguous cells over scales of typically 3 kpc. Hence by studying the \mgii\ absorbers one traces the CGM on kpc scales.

Quasar absorbers observations have estimated the characteristic size of gas clouds to be a few dozen of pc. These results come from indirect photoionisation modelling \citep[i.e.][]{werk14}. Gravitationally lensed quasars more directly constrain the cloud sizes to be less than $\sim$ 30 pc \citep{rauch99, rubin17} but are partially lens-model dependent \citep{lopez18}. One cannot exclude that two independent sightlines hit different metal clouds as expected from density peaks on a single coherent larger structure, the so-called "blobby sheet" model \citep{biggs16,koyamada17}. 

Here we present the serendipitous discovery of extended bright background objects with intervening metal absorbers in a seeing-limited MUSE cube. We probe these absorbers over a total area of $\sim$30 kpc$^2$. This test case illustrates the new information now available to characterise the physical conditions of metal gas clouds. It potentially enables a direct measure of the spatial distribution, clumpiness and metal cloud sizes which, combined with our knowledge of the densities, constrains the cloud gas masses. In fine, it provides a test to the poor-metal mixing scenario by spatially resolving absorption lines on kpc-scales. 

The spatial resolution of the data is key to achieve these scientific goals so that adaptive optics GALACSI MUSE observations will play a crucial role by providing seeing-enhanced observations. In the future, an increased number of targets will be within reach thanks to the collecting area of the next generation of telescopes combine with IFU capabilities such as the HARMONI instrument \citep{thatte16} under construction for the ELT. At higher redshifts, bright targets will be within reach of JWST. 

\vspace{-0.8cm}
\section*{Acknowledgements}
CP is grateful to the ESO and the DFG cluster of excellence `Origin and Structure of the
Universe' for support. RA thanks CNRS and CNES for support for her PhD.

\vspace{-0.7cm}
\bibliographystyle{mnras}
%\bibliography{SQ}
%\bibliographystyle{mn2e}
\bibliography{bibliography.bib}  % see Rapport.bib

% Don't change these lines
%\bsp	% typesetting comments
\label{lastpage}
\end{document}